\documentclass[natbib=false]{article}

\usepackage[letterpaper, margin=1in]{geometry}

\usepackage{fancyhdr}
\usepackage{hyperref}
\hypersetup{
    colorlinks=true,
    linkcolor=blue,
    filecolor=magenta,  
    citecolor=blue,
    urlcolor=blue,
    pdftitle={Leveraging DevOps for Scientific Computing},
    pdfpagemode=FullScreen,
    }

\usepackage[backend=biber, style=numeric-comp]{biblatex}
\usepackage[bitstream-charter]{mathdesign}
\usepackage[T1]{fontenc}

\usepackage{graphicx}
\usepackage{listings}
\usepackage[frozencache, cachedir=.]{minted}

\begin{document}

\title{Leveraging DevOps for Scientific Computing}

\author{Paul Nuyujukian}
\date{Departments of Bioengineering, Neurosurgery, \& Electrical Engineering \\ Stanford Bio-X, Wu Tsai Neurosciences Institute \\Stanford University}

\maketitle

\thispagestyle{empty}

\begin{abstract}
Critical goals of scientific computing are to increase scientific rigor, reproducibility, and transparency while keeping up with ever-increasing computational demands.
This work presents an integrated framework well-suited for data processing and analysis spanning individual, on-premises, and cloud environments.
This framework leverages three well-established DevOps tools: 1) Git repositories linked to 2) CI/CD engines operating on 3) containers.
It supports the full life-cycle of scientific data workflows with minimal friction between stages--including solutions for researchers who generate data.
This is achieved by leveraging a single container that supports local, interactive user sessions and deployment in HPC or Kubernetes clusters.
Combined with Git repositories integrated with CI/CD, this approach enables decentralized data pipelines across multiple, arbitrary computational environments.
This framework has been successfully deployed and validated within our research group, spanning experimental acquisition systems and computational clusters with open-source, purpose-built GitLab CI/CD executors for slurm and Google Kubernetes Engine Autopilot.
Taken together, this framework can increase the rigor, reproducibility, and transparency of compute-dependent scientific research.
\end{abstract}

\section*{Introduction}

Scientific research is becoming ever more increasingly dependent on computational data analysis.
At the same time, data volumes are growing increasingly large, particularly from experimental sources.~\supercite{GilEtAlComputer2007, ReedDongarraACMComm2015}
There are many challenges associated with taming this volume of research data and making it accessible for computational analysis.~\supercite{FillingerEtAlAnalBioanalChem2019,WangEtAlJMedEngTech2020, KhannaEtAlCurrMedImaging2022, GuoEtAlChemComm2022}
Due to the size of the data, a greater fraction of scientific analysis must happen in powerful computing clusters and less on individual, local systems.
This presents novel problems for research domains historically not accustomed to high-performance computing.~\supercite{QuackenbushNatRevGen2001,McClureEtAlNucAciRes2013,AlschnerResMethods2021}
Best practices surrounding data governance, code versioning, and computational library management can help mitigate these challenges, but familiarity with these techniques is limited, particularly for researchers new to cluster computing.~\supercite{WilsonEtAlPLoSCompBIo2017}
Combined with the stark differences between running code locally versus scaling it out with job schedulers, it becomes clear that many researchers are not well-equipped for large-scale scientific computing.~\supercite{DillonEtAlIEEEAdvInf2010,SajidRazaIntConfCloud2013}
Modern researchers need accessible tools that guide and encourage them to adopt best-practices and minimize the challenges they face in conducting computational analyses at all scales and stages of the research lifecycle.

A promising set of solutions to address these needs are the recent development of DevOps tools from the information technology community.~\supercite{EbertEtAlIEEESoftware2016,LeiteEtAlACMCompSurv2019}
Now ubiquitous in industry, these tools have become the new standard for integrated code development and operations deployment, from where the term DevOps originates.
They are most commonly used by teams of web developers to write, test, and deploy Internet applications such as websites and web services.
Fortunately, these sophisticated tools are flexible enough that they may have relevance for scientific computing.
Some elements have already made significant headway into scientific domains--Git repositories for code are considered a best practice.~\supercite{RamSourceCodeBiolMed2013}
Others, like containers, have gained some traction.~\supercite{KurtzerEtAlPLoSOne2017,HaleEtAlCompSciEng2017,SahaEtAlPracExpAdvResComp2018,}
Finally, key components of DevOps frameworks, such as CI/CD engines have had limited adoption.~\supercite{MoutsatsosEtAlSLASDisc2017,KrafczykEtAlPracRepEvalCompSys2019}
Are there opportunities to accelerate scientific research exist with these tools?

In this piece, I argue that core pieces of DevOps frameworks have significant potential to benefit scientific computing.
These benefits extend not just to greater efficiency and productivity (which should be sufficiently motivating in their own right), but also towards important goals like increasing rigor, reproducibility, and transparency of research.
Further, these DevOps components do not act independently, but can be integrated into a unified framework that delivers valuable features to enable scientific research.
At a time when the public trust in science is at historic lows,~\supercite{BurakoffAP2023} and uncertainty regarding scientific integrity can be catastrophic to institutions~\supercite{SaulNYT2023} and fields,~\supercite{PoldrackEtAlNatRevNeuro2017} tools that improve scientific rigor and validity could not be of greater importance.~\supercite{LaineEtAlAnnInternMed2007, PengScience2011}

\section*{Three Pillars of Scientific Computing}

There are three key components borrowed from DevOps tools that make up the scientific workflow framework presented here.
These are 1) containers, 2) Git repositories, and 3) CI/CD engines.
Each will first be discussed independently, followed by subsequent examples of how these components can be integrated in a unified framework to address critical scientific computing needs.

\subsection*{Containers}

One of the largest challenges of any computational research framework is setting up the environment in which the researcher's scientific code will execute.~\supercite{ZakariaEtAlSC22}
This includes elements such as user-level and system-level libraries, programming languages, and data management utilities.
Environment configuration factors can present significant barriers to using multiple platforms, upgrading hardware, or evaluating new or next-generation environments.
Further, library dependency problems can be so daunting that they can require expert knowledge to set up complex computational environments.~\supercite{GamblinEtAlSC2022}
To first-order, the skills necessary to set up and configure a computational environment, which are more in line with system administration, are unrelated to the skills needed to use the environment for scientific research.
For this reason, most computational researchers defer to dedicated, expert high-performance computing or information technology staff to set up and maintain the computational environment.

However, access to these expert staff is limited, and is typically restricted to supporting large computational deployments such as supercomputing environments.
While highly performant, cloud environments often have limited support without expensive support contracts, and even then scientifically-relevant expertise can be limited.
Researchers often have even fewer options for supporting computation on their local laptop and desktop systems, which is where most computational research originates.
For these reasons, local computational environments rarely resemble professionally-managed computational environments, leading to further challenges when attempting to scale out work.

A key tool that can address many of these issues are containers.
Containers are packaged environments with virtualized operating systems that run on host machines.
The most important benefit of a container is that it is portable and static--it operates the same no matter where it is run.~\supercite{BenedicicEtAlISCHPC2019}
If leveraged for scientific computing, they can mitigate challenges associated with configuring multiple heterogeneous environments.
Further, containers are easily distributed, so a single expert, even if remote, can design and share a container relevant to a specific scientific domain, and many can effortlessly benefit from this effort.
Additionally, containers are readily expandable, and thus customizing an existing, well-built container is a straightforward best practice.
Since Open Container Initiative (OCI) containers can run on local systems and also on cluster environments, through apptainer (previously called singularity), they can provide a uniform experience for the researcher regardless of which computational environment they may be using.
Containers are also native to cloud platforms such as Kubernetes, and thus a single portable container can serve virtually every scientific computational environment.
Their uptake has not been as pronounced as some other tools like Git, but containers stand to significantly simplify barriers researchers face when moving between computational environments.

\subsection*{Git Repositories}

The heart of all computational work is the analysis code that performs the data processing and makes the calculations.
This code represents the essence of the scientific data analysis.
In an ideal computational framework, this analysis code is where a scientific researcher should be spending the bulk of their time.
This code generally comprises the foundational ideas reported on in the scientific literature and is the source for most intellectual property arising from computational research.
For these reasons, it is critical that an accurate record of the analysis code be maintained--one that includes the history of its development and changes.
This is critical to maintain both scientific integrity and transparency.

It is generally a well-accepted best practice that code should be kept under version control.~\supercite{VuorreCurleyAdvMethodsPsychSci2018, ByranAmerStat2018}
The most popular version control system in modern practice is Git, which is a distributed version control system.~\supercite{WeberLuoIEEEDataMin2014}
It is ubiquitous in the DevOps community and has a rich ecosystem of repository hosting providers spanning free, commercial, and open-source offerings.
Keeping scientific code under Git version control provides many benefits, including: history tracking, simplified code synchronization across multiple computational environments, and easy collaboration with others.
It is worth noting that for the greatest scientific transparency, researchers should abstain from using \texttt{git rebase} strategies and should only use \texttt{git merge}.

With minimal additional work of integrating it with trusted timestamping (RFC3161/RFC5816), Git repositories can also be leveraged, privately or publicly, as legally defendable records.~\supercite{BuhlmannStartu2021}
RFC3161 integration with Git provides cryptographically secure attestations to a repository's authenticity with respect to the time of commits and integrity with respect its contents.
In cases where the scientific record is challenged, a repository history where revision commits are verifiable can be critical to defending a timeline of work and discovery.
To my knowledge, this methodology has not yet been utilized in any cases surrounding scientific disputes, but I hypothesize would significantly simplify such legal disputes.

A related critical role is for Git repository servers--the web servers that host Git repositories.
These function as the central hub in a decentralized revision environment.
As will be evident in its role in subsequent sections, a Git repository server is the authority for defining and orchestrating workflows.
In this manner, Git can serve as more than just a version control system, but as a key component in scientific pipeline operations.

\subsection*{CI/CD Engine}

The newest and most prominent element of modern DevOps workflows is the CI/CD engine.
Continuous Integration (CI) is the concept of merging incremental code changes into a main branch often.
Continuous Delivery (CD) is the evaluation and deployment of these incremental code changes from CI into a production environment in a largely automated fashion.~\supercite{SinghEtAlIEEECloudComp2019}
These principles originated from the information technology industry and are mostly embodied within DevOps methodology.
Some engines (e.g., GitLab, GitHub, Atlassian) also function as Git repository servers, and can act as a centralized platform for both storing code and executing CI/CD pipelines.
While these tools were designed for the needs of the information technology industry (e.g., revising code and updating a web service in a rapid fashion), many CI/CD engines are well-suited for managing scientific computational workflows.

CI/CD engines act on a repository using a set of instructions (i.e., the pipeline) and execute these instructions in a specified environment.
Various engines use different terminology for the many components comprising a CI/CD platform.
In this work, components will be referred to using the terminology used by the GitLab CI/CD engine.
In GitLab, the environment where a pipeline is executed is termed an executor.
There are various executors, including shell, Docker, and Kubernetes executors.
Custom executors for non-supported environments can also be written.
To support additional environments useful for scientific computing, I have released a slurm executor (\url{http://github.com/bil/gitlab-executor-slurm}) and guidelines on how to successfully deploy a Kubernetes executor on Google Kubernetes Engine Autopilot (\url{https://github.com/bil/gitlab-runner-gke-autopilot}).

Code listing~\ref{ls:basic} shows a simple two-job pipeline written in the format of a GitLab CI/CD YAML pipeline.
Other CI/CD engines have slightly different pipeline definition formats, but have many conceptual similarities.
This simple pipeline has four components: a default section, a stages section, and two jobs.
Elements in the \texttt{default} section apply to all jobs.
The \texttt{image} key defines which Docker container to run jobs in.
In this example, both \texttt{job1} and \texttt{job2} will be run from inside the \texttt{ubuntu:22.04} Docker container.
Both jobs will also run in the environment identified by the tag named \texttt{docker-cluster}.
This environment is a hypothetical computational system registered with the GitLab server that supports running Docker containers.
When the pipeline is executed, the GitLab server signals to the \texttt{docker-cluster} runner process that a pipeline is ready.
The runner process is a small executable that is always running on the computational environment and listening for new jobs to execute.
Technically, the runner process is continually querying the GitLab server for jobs, as the executor may be behind a firewall.
The \texttt{stages} section specifies the sequence of stages to execute.
Individual jobs must belong to a single stage.
All jobs in a common stage execute in parallel.

When signaled, the GitLab runner clones down the repository containing the scientific code that will be executed.
\texttt{job1} runs two processes sequentially.
The first process is a shell script that downloads data onto the local system.
The second process is a Python script that analyzes the downloaded data.
These processes are executed from a working directory at the base of the cloned repository.
When the last process on \texttt{job1} is complete, \texttt{job2} begins and is executed in a similar way.
When this second job is complete, the pipeline ends and GitLab marks the workflow as completed and successful.

\begin{listing}
\begin{minted}{yaml}
default:
  image: ubuntu:22.04
  tags:
    - docker-cluster

stages:
  - stage1
  - stage2

job1:
  stage: stage1
  script:
    - sh ./download-data.sh
    - python3 analyze-data-step1.py

job2:
  stage: stage2
  script:
    - python3 analyze-data-step2.py
\end{minted}
\caption{A two-job pipeline in GitLab CI/CD YAML format.}
\label{ls:basic}
\end{listing}

CI/CD engines support a myriad of features for designing complex workflows.
These include conditional logic, parallelization, dependencies, and artifact generation.
In many scientific workflows, the precise workflow cannot be statically enumerated and is instead dynamically assembled (e.g., dependent on aspects of the specific dataset processed or a response from a web query).
For these cases, CI/CD engines support dynamic pipeline definitions, where sections of the pipeline can be defined based on the output of code executed at the appropriate time.

The benefits of CI/CD engines to scientific computing are significant, as they bring structure to an otherwise challenging problem.
Pipelines increase research efficiency because they simplify the challenges with iteration.~\supercite{BiswasEtAlIntConfSofEng2022,DjaffardjyEtAlCompStrucBiotech2023}
When parameters are exposed as variables into the pipeline appropriately, it becomes straightforward to perform validation across large datasets.
Further, pipeline development naturally leads to reproducibility, as scientific analysis can be rerun easily across multiple datasets to ensure consistent results.
They ease challenges with collaboration and sharing, as the inputs and outputs are well defined.
In a functioning pipeline, provided the input data and the Git repository, another researcher should be able to execute the pipeline and validate the results.
These benefits are tightly aligned with increasing scientific rigor, reproducibility, and transparency.
CI/CD engines, while not designed for scientific workflows, are a natural solution to the growing complexity of scientific computing and complex workflows.

\section*{Integration Examples}

In this section, three examples will be discussed that integrate the three pillars discussed above.
Each demonstrates how the tools from DevOps can further scientific computing.

\subsection*{Centralized Pipelines}

\begin{figure}
    \includegraphics[width=\columnwidth]{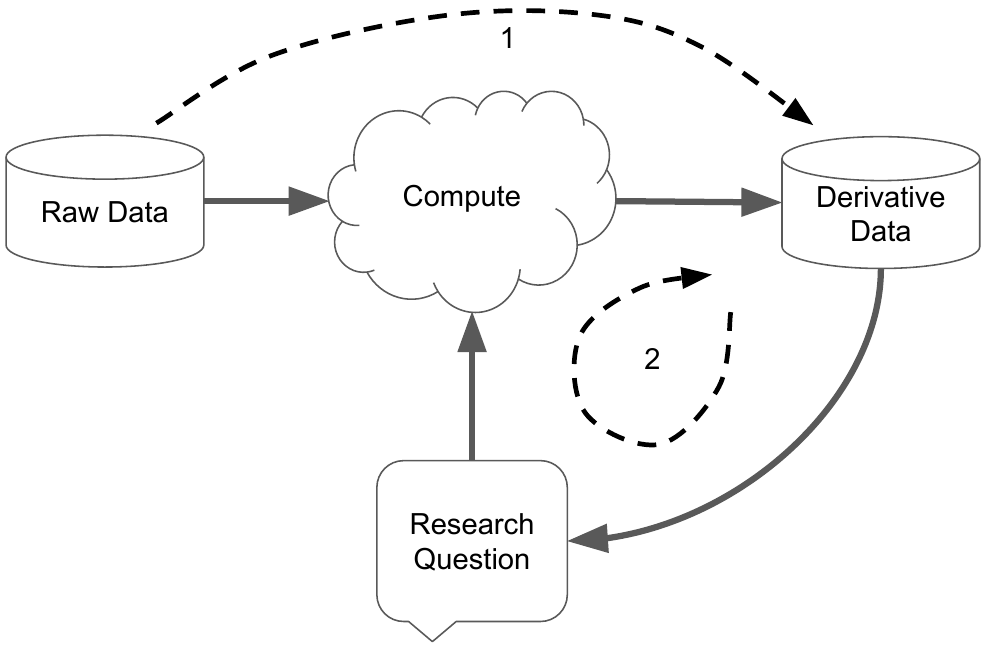}
    \caption{Diagram of centralized pipeline.
    All data processing takes place within a single computational environment, often under the governance of a job scheduler.
    Dashed line 1 represents the pipeline that converts raw data to derivative data.
    Dashed line 2 represents the pipeline that investigates a research question and leads to scientific insight by generating additional derived data.
    Solid black lines represent flow of information and/or researcher effort.}
    \label{fig:centralized}
\end{figure}

Centralized pipelines are the most familiar form of scientific computing.
These pipelines are scientific workflows that execute in a single, homogeneous environment.
An example centralized pipeline appears in Figure~\ref{fig:centralized}.

Centralized pipelines are executed in environments traditionally and most commonly used for high-performance computing, including supercomputing and shared computing clusters.
These environments are carefully managed by expert staff where both hardware and software are actively managed.
Most commonly, some kind of job scheduler (e.g., slurm, SGE) is utilized to coordinate resources in an organized way.
Containers in these environments are most commonly supported through apptainer.

A second, and newer, environment for centralized pipelines is cloud computing.
This can include staff-managed virtual machines, but is more commonly shifting to serverless, containerized frameworks such as Kubernetes.
Here again, job schedulers exist, and span cloud vendor specific solutions to recent open source offerings (e.g., Apache Beam).

Scientific computing workflows in centralized pipelines perform data processing.
In a typical workflow, input data is copied into the computing environment, and processed to generate derivative data.
This derivative data may be further processed in subsequent pipelines.
The defining feature of a centralized pipeline is that all automated elements of the workflow occur within the same environment.

A major advantage of the proposed framework advanced in this work is that by using CI/CD engines, workflows can be largely agnostic to the specifics of the environment they are executed in.
This is possible due to the tight integration of the three DevOps elements discussed above.
Since CI/CD engines like GitLab support custom executors, scheduler management can be abstracted away to the executor and scheduler parameters can be directly specified in the centralized workflow.
Similarly, parameters for other environments can be specified in the same workflow, and environments can be swapped around by modifying the target environment the pipeline should be executed in.

Code listing~\ref{ls:central} illustrates this idea.
This is the same workflow as in Code listing~\ref{ls:basic}, now augmented with resource specifications for two computational environments.
As written, this workflow will execute in the \texttt{slurm-cluster}, as defined by the \texttt{tags} in the \texttt{default} section.
The GitLab custom executor for slurm written by the author looks for the \texttt{SLURM\_PARAMETERS} variable associated with each job and passes those parameters onto the call to \texttt{sbatch} and executes the container with singularity (reference omitted due to double-blind review).
In this example, \texttt{job2} is more computationally intensive than \texttt{job1}.

\begin{listing}

\begin{minted}{yaml}
default:
  image: ubuntu:22.04
  tags:
    - slurm-cluster

stages:
  - stage1
  - stage2

job1:
  variables:
    SLURM_PARAMETERS: -c 1 --mem 2G -t 1:0:0
    KUBERNETES_CPU_REQUEST: 1
    KUBERNETES_MEMORY_REQUEST: 2G
  stage: stage1
  script:
    - sh ./download-data.sh
    - python3 analyze-data-step1.py

job2:
  variables:
    SLURM_PARAMETERS: -c 5 --mem 40G -t 5:0:0
    KUBERNETES_CPU_REQUEST: 5
    KUBERNETES_MEMORY_REQUEST: 40G
  stage: stage2
  script:
    - python3 analyze-data-step2.py
\end{minted}
\caption{A two-job, centralized pipeline with resource specifications for both slurm and Kubernetes environments.}
\label{ls:central}
\end{listing}

Further, in this example, the only change the researcher would need to make to execute this pipeline in a Kubernetes environment would be to switch the value for \texttt{tags} to the value of the Kubernetes executor.
This is possible because: 1) the scientific code being run is stored in a central Git repository server, 2) the libraries that the code depends on are assembled in a container accessible to both environments, and 3) the CI/CD engine has registered executors defined for both computational environments.
This example assumes that the work of registering these slurm and Kubernetes executors to the GitLab server has been previously performed.
The Kubernetes executor ignores the value of the \texttt{SLURM\_PARAMETERS} variable, but instead respects the values of the Kubernetes variables for each job to allocate CPU and memory resources.
Time is not typically necessary to specify for Kubernetes as pods are allocated dynamically for each job and released when a job is complete.
A recent advantage of Google Kubernetes Engine is the Autopilot cluster mode, where node pools are not defined, simplifying both cluster administration and billing.
The author released guidelines on how to deploy a GitLab Kubernetes executor for this computational environment, with the necessary adjustments to both the helm configuration and the cluster to support this mode of operation.

As demonstrated in this integration example, centralized pipelines are well supported by DevOps tools.
By specifying pipelines in a common format known to the CI/CD engine, scientific workflows can become largely agnostic to the specific computational environment they are executing in and, provided that an appropriate executor exists, easily pivoted to other environments.

\subsection*{Decentralized Pipelines}

\begin{figure}
    \includegraphics[width=\columnwidth]{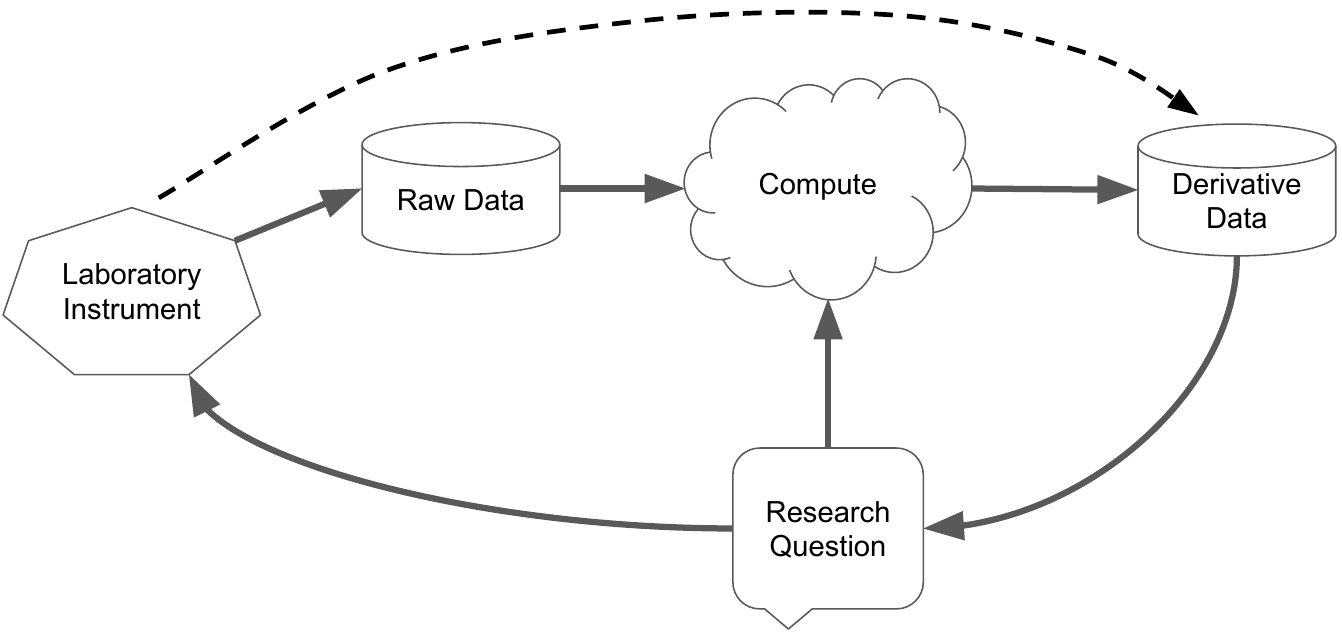}
    \caption{Diagram of decentralized pipeline.
    Here, data processing does not necessarily take place within a single computational environment.
    The dashed arrow represents a key pipeline that spans the laboratory instrument and the computational cluster.
    In this case, the pipeline manager must be capable of managing heterogeneous environments.
    CI/CD engines are well suited for this task.
    Solid black lines depict flow of information and/or researcher effort.}
    \label{fig:decentralized}
\end{figure}

Unlike centralized pipelines, decentralized pipelines do not execute in a single computational environment.
Instead, they execute in heterogeneous environments across local systems, scientific instruments, supercomputing clusters, and cloud environments.
The nature of this heterogeneity makes them more complex to manage, test, and validate.
While these environments may be less common among classical scientific computationalists, they are ubiquitous among experimental researchers.
As the needs of experimentalists are becoming more computing-dependent, the demand to operate decentralized pipelines will continue to grow.
A diagram of a decentralized pipeline appears in Figure~\ref{fig:decentralized}.
Not only do these experimentally-originating pipelines answer scientific research questions, but, commonly, their findings spark new questions that require returning to the laboratory instrument to answer.

A unique challenge posed by decentralized pipelines is that there are limited solutions in place for this growing need.
High-performance computing job schedulers do not typically extend to scientific instruments in a research lab.
Similarly, cloud computing vendor offerings for cloud pipelines primarily operate within the walled gardens of the vendor's cloud.
Here again, tools from DevOps can enable critical functionality and deliver solutions to researchers looking to deploy decentralized pipelines.

\begin{listing}
\begin{minted}{yaml}
default:
  image: ubuntu:22.04

stages:
  - stage0
  - stage1
  - stage2

job0:
  tags:
    - scientific-instrument
  stage: stage0
  script:
    - powershell ./upload-data.bat

job1:
  tags:
    - slurm-cluster
  variables:
    SLURM_PARAMETERS: -c 1 --mem 2G -t 1:0:0
    KUBERNETES_CPU_REQUEST: 1
    KUBERNETES_MEMORY_REQUEST: 2G
  stage: stage1
  script:
    - sh ./download-data.sh
    - python3 analyze-data-step1.py

job2:
  tags:
    - kubernetes-cluster
  variables:
    SLURM_PARAMETERS: -c 5 --mem 40G -t 5:0:0
    KUBERNETES_CPU_REQUEST: 5
    KUBERNETES_MEMORY_REQUEST: 40G
  stage: stage2
  script:
    - python3 analyze-data-step2.py
\end{minted}
\caption{A three-job, decentralized pipeline that executes on a laboratory computer acquiring data from a scientific instrument, a slurm cluster, and a Kubernetes cluster.}
\label{ls:decentralized}
\end{listing}

Code listing~\ref{ls:decentralized} describes a three-job decentralized pipeline that exemplifies a common use case for an experimental researcher acquiring data.
Once the experiment is finished and data acquisition from the scientific instrument is complete, it must be uploaded to data storage (e.g., on-premises or cloud) and then processed to make it suitable for analysis.
The computer that manages the scientific instrument is rarely capable enough to perform any heavy computation on this acquired data.
For heavily-used instruments like in shared facilities, there is also little downtime and data must be quickly moved off the limited storage space of the instrument system.
Thus, data processing must occur in some other environment, like a computing cluster or the cloud.
In this example, a new job, \texttt{job0} is introduced which runs on the scientific instrument.
In GitLab, this would commonly be done with the shell executor, which supports Windows command line, Mac shell, and Linux shell scripts.
Windows support here is crucial as many instrument vendors only provide software running on Windows to operate their expensive equipment.
Here, the Docker image would be ignored as the shell executor does not support Docker, and the script would run in the Windows shell.
Technically, the \texttt{powershell} line does not need to be specified, but it is provided here for illustrative purposes to emphasize that \texttt{job0} is running on Windows.
Once \texttt{job0} is complete and the data has been uploaded, \texttt{job1} executes on the slurm cluster, followed by \texttt{job2} on the Kubernetes cluster.

This example demonstrates how DevOps tools can deliver novel computational workflows that meet researchers' needs to span environments.
This is particularly of significant relevance to experimental researchers who generate increasingly larger volumes of data from their laboratories and are in need of workflows that can cross their heterogeneous spaces with indivdiual workflows.

\subsection{Minimizing Scale Out Friction}

\begin{figure}
    \includegraphics[width=\columnwidth]{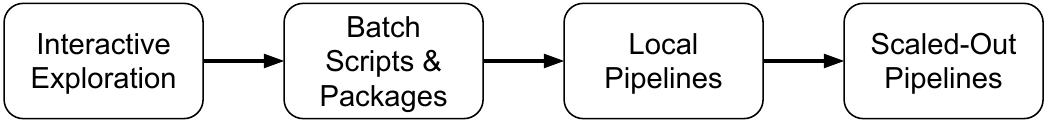}
    \caption{Diagram of progression of scientific computational work from early local exploration on a laptop or desktop to eventual scale out on a computational cluster.}
    \label{fig:scaling}
\end{figure}

An old and familiar challenge for computational researchers is the problem of scaling out an analysis.~\supercite{ChenEtAlSoftPracExp2017}
A common pattern of computational research inquiry typically starts with interactive exploration of a small portion of the dataset.
This often involves plotting and visualization in an interactive fashion, usually on a local system like a laptop or desktop.
This can include interactive command line shells or web-based notebooks.
The scientific code written during this exploratory phase changes heavily while the researcher is looking to understand the nature of the data and what analyses may be appropriate.

If a particular analysis is found to be fruitful, they may cross-check the result with another small portion of the dataset.
However, inevitably, to validate the analysis, the scientific code must be run on all, or a significant portion, of the dataset.
This process of scaling out the analysis can often be daunting, particularly for researchers new to computational methods.
While they may have the skills to perform the analysis locally, the prospect of modifying their code to operate on a supercomputing cluster and against a job scheduler can appear daunting.
Further, as they move to a new computational environment, they may encounter new library or dependency problems, some of which they may be poorly equipped, or lack appropriate system permissions, to resolve.
Even with familiarity, the process of scaling out can be time intensive if a new computational environment must be set up in the cluster.
These hurdles can present challenges that slow scientific productivity, and in extreme cases, result in abandoned and incomplete investigations.

Once again, tools from DevOps can be helpful for researchers in minimizing the challenges they face in scaling out analyses found on small-scale interactive exploration.
A diagram of this progressive methodology appears in Figure~\ref{fig:scaling}.
This is accomplished by utilizing a common Docker container for every stage of the computational scientific investigation.
This container needs to have libraries that support interactive exploration as well as those utilized at the scale out stage by the computational cluster.
So long as the researcher always conducts their interactive work within the container, then the process to scaling out the work is minimized.
The author maintains such a container for their research group, primarily supporting Python-based computation: \url{http://github.com/bil/comp-env}.
This container has BLAS/LAPACK optimized versions of numpy and scipy, HDF5, and a number of other common numerical Python libraries.
In addition to iPython, it also includes JupyterLab and Visual Studio Code, which are run as local web servers.

Researchers can use web notebooks and the vscode web interface to perform interactive work that supports visualization and writing code.
As code matures and is ready to move from notebooks to batch scripts, researchers can use the same container to rapidly test out their assembled Python scripts and packages.
Once their code works in a batch script format, they can begin to assemble their GitLab CI/CD pipeline, and test iteratively to validate it using the Docker executor on their local system.
This permits them to validate pipeline elements with limited parallelism in a fashion identical to what would be experienced in the cluster environment, while not needing to wait in the job scheduler queue.
Once their pipeline is nearly validated, they can switch the \texttt{tags} value and direct their pipeline to the cluster environment for scale out.
The experience is straightforward and involves incremental steps to scaling out work that are all executed in the same system library environment of the container, minimizing barriers commonly faced.

If adhered to, the guiding principles of DevOps stand to significantly simplify the challenges faced by computational researchers as they transition work from the proof-of-concept stage on small portions of their datasets to the larger-scale environments of high-performance computing against their large datasets.

\section*{Beneficial Side Effects}

While the above three integration examples highlight the opportunities DevOps tools can directly bring to scientific computation, there are additional beneficial side effects.
Three such examples will be discussed below.

\subsection*{Increased Rigor, Reproducibility, and Transparency}

A key benefit of employing these DevOps tools is in the increased rigor, reproducibility, and transparency of scientific research.
These tools help create foundational platforms for research.
They can facilitate scaling and parallelization, increasing a single researcher's efficiency and productivity.
This means more time can be spent validating hypotheses and analyses to ensure scientific claims are well defended.
Foundational platforms such as these help raise the bar for the levels of validation considered acceptable for publication within a field.
The more ubiquitous tools such as these become within a discipline, the higher the quality of rigor becomes.

At the same time, these tools permit easier reproducibility of research findings.
The use of containers eliminates the need for other researchers to configure their own libraries to reproduce a result.
Git repositories housing the scientific analysis code, the pipelines for processing the data, and details on the resource needs of each job in the pipeline; significantly lower the barrier to reproducibility.
Finally, all these tools, if properly shared and disclosed, improve the transparency of the work being performed.
In an ideal world, a well-structured data pipeline should process raw data to derivative data, and include the specific pipelines that generated each figure in a scientific manuscript.
Current scientific standards do not currently require that level of transparency, but these tools present a viable path towards such a reality.

\subsection*{Facilitating FAIR Principles and Interoperable APIs}

FAIR principles are a set of guidelines oriented towards improving research data reuse.~\supercite{WilkinsonEtAlSciData2016}
The DevOps tools described here also contribute in important ways to enabling FAIR principles.
Data processing pipelines encourage the practice of a single, common pipeline for as broad a category of data as possible, with rich metadata informing the pipeline on how data should be processed.
This requires datasets to have unique identifiers with informative metadata, making them well-suited to be indexed.
As mentioned in the prior section on reproducibility and transparency, DevOps frameworks make data accessibility a central feature.
This goes hand in hand with reusability.
Pipelines also facilitate interoperability.
Slightly different file formats, metadata, or other elements are readily supported with small adjustments to the common pipelines.
Reprocessing the datasets under the revised pipeline make the data compatible with new formats in an automated fashion with limited manual intervention.

An important benefit that DevOps frameworks bring to scientific research and computing is the ease with which interoperable APIs can be developed.
While significant focus has been placed on common file formats across various domains of computing, a perhaps underappreciated opportunity is in the development of common APIs.
Major strides forward in computational domains have been made when a common API emerges.
This allows data providers to build towards that API.
It also permits data consumers to build frameworks based on that API.
This has been true of popular APIs both old and recent, such as REST APIs with HTTP, SQL for databases, domain-specific Python libraries such as neo for neuroelectrophysiology, and, most recently, the Hugging Face libraries for machine learning.
The DevOps tools described here can help facilitate the emergence of these common APIs across many domains of scientific research, which would rapidly accelerate domain progress.

\subsection*{Greater Compliance with Open Data Mandates}

A third area DevOps tools may critically contribute is in compliance with open data mandates.~\supercite{EswarappaNature2022}
These mandates are being placed on scientific research at nearly every level, from funding organizations, publishers, and even government agencies.
For years, many scientific journals have employed open data requirements as conditions of publication.
Similarly, funding agencies spanning public entities (e.g., US National Institutes of Health) to private ones (e.g., Howard Hughes Medical Institute) have adopted open data requirements coinciding with publication.~\supercite{NIHPublicAccess2008,DownsUSAIN2016,HHMIOpen2020}
Most recently, the US executive branch mandated that all scientific publications funded by Federal sources must make their data public at the time of publication.~\supercite{BrainardKaiserScience2022}
These trends from many elements involved with scientific research point to an undeniable trend--research data must be public.

However, many researchers are not well-equipped to easily comply with these mandates.~\supercite{KaiserBrainardScience2023}
Further, it may be incomplete to simply require open data mandates without additional guidance on how data should be released.
Releasing functionally inaccessible data publicly is no better than keeping data private.
As in prior sections surrounding scientific transparency and FAIR principles, DevOps tools, if employed well, stand to significantly increase compliance with the \textit{spirit}, and not just the letter, of these open data mandates.

\section*{Related Efforts}

To the author's knowledge, this is the first framework that supports decentralized pipelines that span multiple heterogeneous environments for scientific purposes.
There are some similar efforts to the work presented here that are relevant to discuss.
Jacamar CI is a project looking to deploy a custom GitLab executor for HPC environments supporting a number of job schedulers.
A key distinction is that Jacamar is designed to be deployed by HPC administrators and not by the researcher.
As such, there is limited control over where and how the GitLab runner is registered.
This could present limitations in deploying decentralized pipelines.

Similarly, the Joint Laboratory for Extreme Scale Computing (JLESC) has also been evaluating the applicability of CI tools for HPC environments.
Thus far, their work has been focused on integrations between GitHub and GitLab.
In a related manner, the National High-Performance Computing Center at the Karlsruhe Institute of Technology encourages the use of GitLab CI in their high-performance computing clusters.

\section*{Limitations}

The framework presented in this work is not without limitations.
The first is to admit that these tools are relatively novel and knowledge about how to use them among scientific communities may be lacking.
This presents a challenge with respect to uptake and adoption because the educational frameworks may not be very present.
Second, these tools involve slightly more work than running directly on one's local system when prototyping code.
The temptation to do what is ``quick'' when prototyping individually is significant, and can present a hurdle when attempting to scale out work because a computational container was not used from the outset.

Another limitation is the increased burden on developers for maintaining infrastructure.
There is work in maintaining a reference container and keeping it up to date with the newest libraries.
Ensuring that custom executors are functional requires developer time and resources.
These additional needs can be challenging to meet on limited academic budgets and require planning and prioritization.

\section*{Conclusion}

This work describes three key components of DevOps frameworks and how they may be useful in improving scientific computing.
These three elements are containers, Git repository servers, and CI/CD engines.
Containers address the problem of library and dependency challenges.
Git repository servers serve as centralized stores for scientific code.
CI/CD engines leverage the prior two elements to deploy pipelines onto computational environments spanning local systems, supercomputing clusters, and cloud environments.
Each of these components neatly integrates with the other, resulting in a combination that is greater than the sum of its parts.
Three integration applications were presented: centralized pipelines, decentralized pipelines, and minimizing scale out friction.
Finally, three side benefits to DevOps tools were discussed: increasing rigor, reproducibility, and transparency; facilitating FAIR principles and interoperable APIs; and improving compliance with open data mandates.
Taken together, these examples and arguments suggest that scientific computational research stands to benefit significantly by borrowing from the extensive investment into DevOps tools made by the IT community.
Adoption of these tools by the various scientific fields is expected to further overall scientific productivity.

\printbibliography

\end{document}